# Ontologia para monitorar a deficiência mental em seus défícts no processamento da informação por declínio cognitivo e evitar agressões psicológicas e físicas em ambientes educacionais com ajuda da I.A*

Cuidado:Este conteúdo cita linguagem inapropriada de outros.

Bruna Araújo de Castro Oliveira [1*]
Instituto Federal do Espírito Santo /IFES-Serra

**Abstract**

A intenção deste artigo é propor o uso da inteligencia artificial para detectar através de análise por ontologia UFO o surgimento de agressões verbais e físicas relacionadas a deficiências psicossociais e seus agentes provocadores, na tentativa de se prevenir consequências catastróficas dentro de ambientes escolares.

## 1. Introdução

A deficiência psicossocial é uma realidade em ambientes educacionais e não pode ser simplesmente negligenciada ou desconsiderada visando a evasão escolar destes usuários em favor de outros não deficientes mentais, o que contraria a função social em instituições de ensino: A obrigação destas e seus funcionários é identificar as pessoas responsáveis primeiramente,detê-las e encontrar soluções que atendam as necessidades dos usuários com problemas mentais no intuito de fornecer um suporte a sua manutenção nestes ambientes artificiais de obtenção de conhecimento,visto que pesquisas mostram que pessoas neste estado mantidas em ambientes educacionais e consequentemente obtendo uma maior escolaridade possui menores chances de desenvolver fala esquizofrênica...Em estudos recentes realizados (Elite dos Santos et al,2014) mostram que as dificuldades de comunicação estão associadas a diversas patologias…...Há evidencias da correlação entre transtornos mentais e alterações nas habilidades cognitivas,comunicativas e linguísticas. Essas alterações podem envolver principalmente quatro processamentos cognitivos: discursivo,pragmático,léxico semântico e prosódico,nos níveis compreensivo e expressivo.”, o que com certeza prejudica o desenvolvimento da aprendizagem deste aluno e que foge ao objetivos de uma instituição realmente comprometida com a educação.

Primeiramente neste artigo será evidenciado o problema através da análise de um modelo de ontologia onto-Uml para identificar as causas do possível surgimento da “fala esquizofrênica” nos alunos deficientes para se chegar então ao objetivo de outro modelo a ser proposto para prevenção e quem sabe uma possível reabilitação para convivência em uma futura sociedade utópica onde haja “respeito as diferenças e igualdade para todos”.
Este será apenas um dos passos para obter o resultado do desenvolvimento de uma cultura de inclusão social programada com uso de ontologias de rastreamento das origens e eliminação de suas suas causas,suporte e reabilitação de pessoas que possuem o mesmo direito de estudar como os outros e que não devem ser banidas do sistema educacional devido as suas divergências,possíveis deficit intelectuais intrínsecos e os preconceitos que advêm dessas,através do uso de inteligencia artificial.”Se a inclusão posiciona -se de uma forma contrária aos movimentos de homogenização e normalização “ (Sassaki,1999) então que seja proposto um modelo de “heterogeneização” e “infornormalização” da inclusão.

## 2. Background

### 2.1. Deficiência Mental

Do processo da **constituição do sujeito de doença mental** (Foucault,1954/1984): ” A patologia mental é encarada pelo indivíduo como elementos que se fecham sobre si mesmo,criando um mundo autônomo,em contraponto com à realidade normal do indivíduo. Isto ocorre nas alucinações e nos delírios. Em caso extremo,na esquizofrenia,por exemplo,”o doente é absorvido pelo mundo da doença”….
Focault ressalta a ponte entre história individual e história cultural social. A patologia de determinada história psicológica e individual não deve ser reduzida aos fenômenos restritos da existência e da percepção personalista do sujeito numa ontologia separada da história e da cultura.” A doença tem sua realidade e seu valor de doença apenas no seio de uma cultura que a reconhece como tal”.Nesse aspecto convém ressaltar que o ambiente cultural social participa intensamente do processo que criação deste mundo patológico,e em outras palavras o mundo ao nosso redor colabora para que se potencializasse a patologia a um nível crítico.
Em síntese uma doença mental que se encontra numa forma branda ,imperceptível ou pouco perceptível pode ser agravada em um nível crítico em ambientes que contribuam para esta situação através de agentes físicos com intenções de segregação social através de discursos de ódio. Destaca-se aqui que quem cria o ambiente hostil como um agente social são as pessoas ,agentes físicos que fazem que o lugar seja propício para incentivar os sintomas negativos dos transtornos mentais.

Deste modo atualmente as deficiências não são apenas vistas como incapacidades funcionais orgânicas,mas também como às limitações no desenvolvimento de atividades e as restrições de participação da vida social.

### 2.2. Memorial: Entendendo uma “ontologia do acidente”,de onde surge o trauma?

Para uma simples percepção inicial coloca-se em voga a brutalidade específica de uma ontologia da deficiência oriunda de uma ontologia do acidente,baseando -se nas colocações de Caterine Malabou para dar visibilidade sobre a peculiaridade de um “relato memorial” e diferenciar a deficiência psicossocial de outras deficiências ,numa relação ontológica dos acidentes produtores ou acentuadores da deficiência e seus efeitos no contexto educacional.

Segundo Malabou a “Biografia dos sujeitos é rompida por acidentes que é “impossível se reapropriar da palavra ou pela rememoração” pois os mesmos não tem significação,mas seus efeitos mudam o sentido de uma forma singular de vida.

“ A destruição permanece um acidente quando devia ao contrário ser considerada como uma espécie de acidente,jogo de palavras que visa dizer que o acidente é uma propriedade da espécie,que a capacidade de transformar por efeito da destruição é um possível,uma estrutura existencial...É por isso que reconhecer a ontologia do acidente é uma tarefa filosoficamente difícil: É admiti-la como uma lei, ao mesmo tempo lógica e biológica,mas como uma lei que não permite antecipar nada sobre seus próprios casos. Uma lei surpreendida pelos seus próprios casos. A destruição ,por princípio,não responde a sua própria necessidade ,não confirma,quando acontece,sua própria possibilidade. (MALABOU,2014,p. 30).

Este fato filosófico demonstra a finalidade da ontologia deste artigo,mostrar de como uma destruição social de um “euser” arquitetada aprimora um sujeito deficiente para transforma lo num ser aparte em prol de elevação de outros numa visão capacitista de cultura na qual vivemos. O que resta então para o ser destruído :Repetir a destruição que lhe foi imposta pelo “acidente”?...porém esta questão não caberá ser respondida aqui especificamente,deixando-se para trabalhos posteriores .

**3. A utilização de modelos Onto-uml e UFO para desenvolvimento do caso de domínio da “Antologia do acidente” para análise de dados:**

Nas palavras de Studer et al.,uma ontologia é uma especificação formal e explícita de uma conceituação compartilhada .É uma teoria lógica utilizada para capturar os modelos pretendidos de uma conceituação e excluir os não pretendidos ,ou seja: Uma teoria utilizada para especificar e explicitar uma conceituação.
]
A UFO fornece uma teoria de distinções ontológicas que aborda uma tipologia de universais como os tipos sortais como os tipos sortais rígidos(kind,subkind),sortais anti rígidos(phase,role) e os tipos dispersivos não sortais (que definem conceitos mais genéricos),tais como Mixin,category,rolemixin)entre outros,juntamente com algumas restrições para regular a construção de modelos ontologicamente consistentes. A fim de assegurar uma formalização adequada,esta ontologia baseia se no módulo UFO-C de entidades sociais que possuem aspectos da UFO-A e UFO-B.Os fragmentos de UFO utilizados neste contexto de são:Category,Agent,Event,kind,phase,Mode.

Segundo em artigo de Silva et al(2020)o suporte provido pela ontologia para abstrair a origem dos dados e possibilitar a realização das predições ocorre dentro da abordagem. É a partir dela que os dados podem ser padronizados em repositórios usando o mesmo schema para posterior geração e aplicação de modelos de Machine learning:Assim sendo mais adiante nas discussões e propostas sugere-se a utilização de um método para aplicação de algorítimos de Machine Learning para criação de base de dados mais robusta.

### 3.1 Início do ciclo:Da análise conceitual do "Acidente causador " para se chegar a um modelo de predição

Na origem do ciclo numa prévia análise básica podemos identificar as fontes de dados para definir o modo de acesso e captura dos dados do agente causador através da modelagem da antologia de referencia básica conforme figura 1 abaixo.

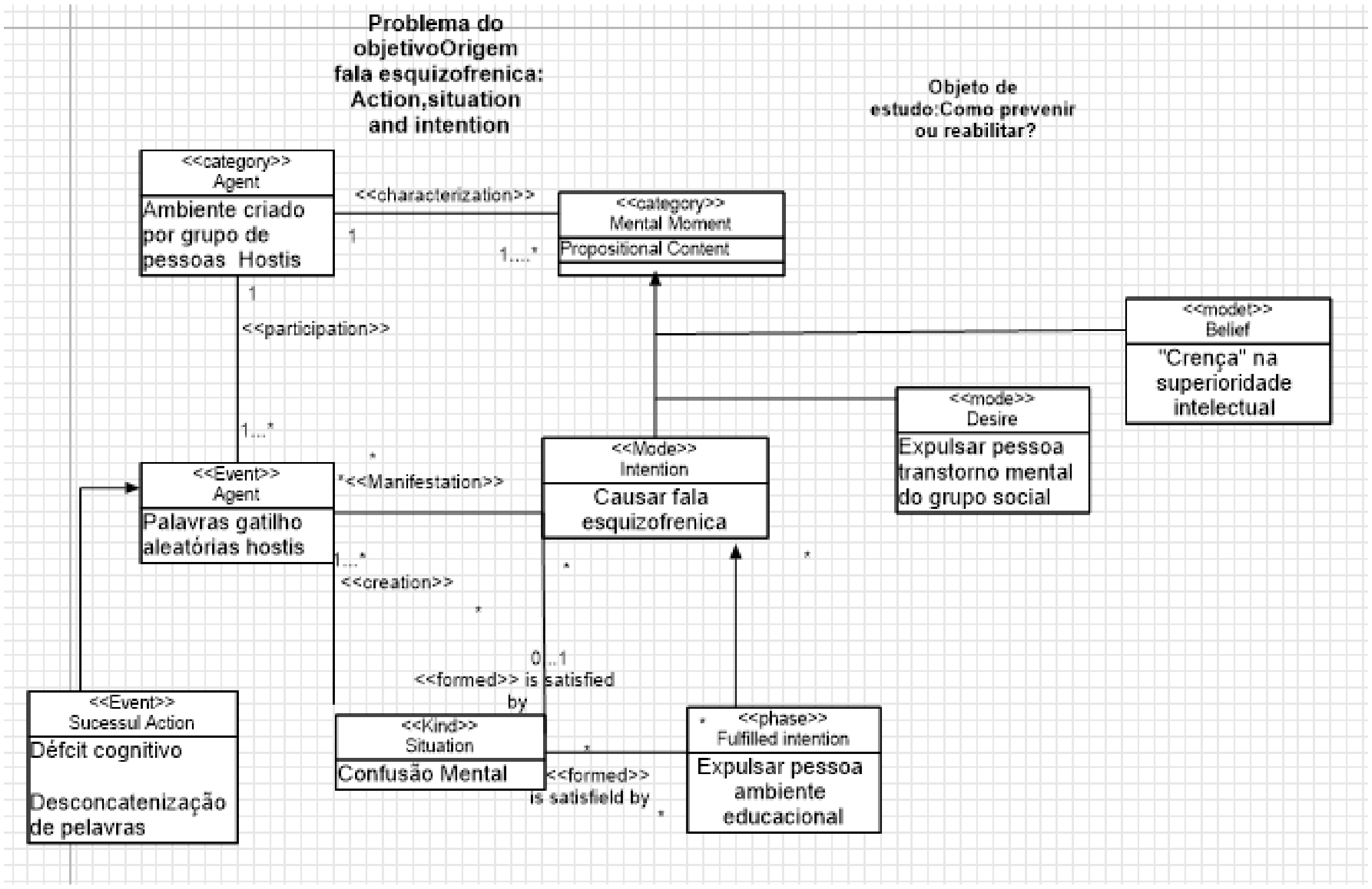

Figura 1:Início do ciclo:Modelo onto-uml básico de referencia do surgimento da fala esquizofrênica.

Neste modelo usa-se principalmente fragmentos de modulo UFO-C de entidade social de módulo substantials afim de colocar um memorial de uma fase de vida do usuário deficiente para representar as circunstâncias do domínio necessárias para possibilitar a identificação de padrões,e um melhor entendimento dos fatores que influenciam o início do desenvolvimento das características negativas de uma doença mental, e o modus operanti dos agentes causadores.

No módulo **Intentional Moment** o principal conceito é **Mental Moment** que engloba 3 mode:Intention,Desire e Bielief que representam o propósito que almeja o agente causador oriundo de seus "sentimentos" sobre o aluno deficiente mental que sofre o efeito dessa causa. O agente causador acredita que possui superioridade intelectual e que deva expulsar a pessoa que não se ade que ao nível  deste grupo do ambiente educacional a fim de privá-lo de um conhecimento por não “ser merecedor”

No módulo **Intention** composto por **situation** e **fullfiled intention** vemos que a relação entre esses conceitos encontra-se na intenção de causar a fala esquizofrênica através de uma confusão mental para especificamente expulsar o aluno deficiente do ambiente educacional.

Na modulo Agente temos 2 eventos associados: **Agent** e **sucesfull action** ,onde o agente é considerado o ambiente criado por alunos não deficientes que proferem palavras gatilho aleatórias hostis com o intuito de se levar a um défict cognitivo e desconcatenização de palavras e criar uma situação de confusão mental no aluno com transtornos mentais.

### 3.2. Evolução do Modelo :Distinção dos Agentes Social e físico especificando sua intenção e ação

Num desenvolvimento mais refinado desta ontologia **complementando os módulos já explicados acima** pode se colocar um novo módulo kind denominado pessoa composto por roles estudante deficiente e não deficiente para diferenciar  o agente causador daquele que sofre,onde o módulo Intentional Moment se aplica ao agente não deficiente, o mode intenção se especifica numa intenção hostil de provocar a fala esquizofrênica,ridicularizar e humilhar o aluno deficiente através de uma confusão mental criada pela ação de provocar um défict cognitivo se especializando em desconcatenização de ideias através de um discurso indireto provocativo com uso de bad setences num discurso de ódio que explica a origem de uma cadeia de ações e possíveis reações de quem sofre,que evidencia quem é o agente físico que inicia uma ação que gera um encadeamento de eventos recursivos em um ambiente hostil especificando de um tipo instituição como agente agente social conforme figura 2 abaixo,que podem levar a uma disposition de Agressividade física.

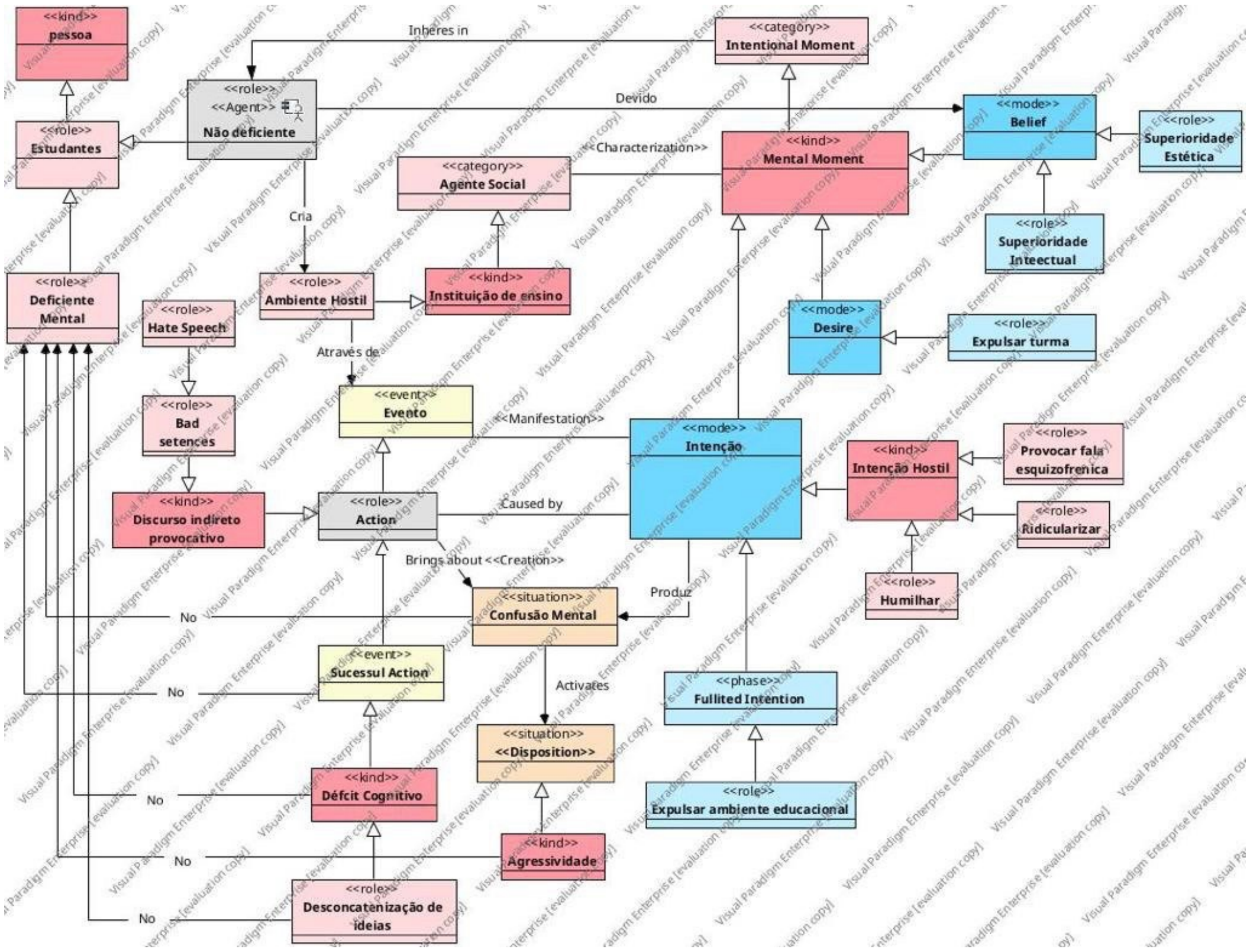

Figura 2 : Especificando os agentes e a disposition de reação negativa do aluno que é alvo de ataques.

O cenário que explica esta ontologia de uma forma mais sucinta e geral pode ser representado num ambiente escolar onde o aluno deficiente mental que **não possui bom desempenho** em sala de aula e talvez sequer conhecimento de sua doença é ridicularizado,humilhado,marginalizado usando sua maneira divergente de agir e se expressar como meio propício para este intuito,sempre ouvindo as mesmas palavras ofensivas,referindo -se as suas atitudes e maneiras,gostos e aparência física, em grande parte de modo indireto mas específico,a ponto de se chegar a um stress emocional muito forte manifestado dependendo da disposition em shutdowns e mutismo,ou provocar uma reação de resposta com fuga de ideias ou ideias desconcatenadas,numa confusão mental que piora ainda mais seu desempenho escolar,podendo chegar a uma agressividade física.

E mais especificamente, usando um cenário memorial como relato de que se trata a ontologia criada,Cita-se: Estudante de graduação em universidade pública com diagnóstico de esquizofrenia com desempenho acadêmico de mediano para pior ouve comentários dentro da sala de aula e corredores tais como:"Burra esquizofrênica","esquizofrênica horrorosa","vamos expulsar esta burra horrorosa da ".FES","essa doida baranga não levanta nem um pau mais","não adianta nem tentar usar suas muxibas(referente aos seios)","não sabe nem ler a burra","completamente maluca","burra doida e horrorosa","não sabe nada de química"(pois reprovou em química orgânica 2),"foi fácil provar que essa horrorosa não passa de uma burra","sabe nada de computador","burra da internet"(pois possui graduação em análise de sistemas),além de outras palavras de baixo calão de cunho sexual. Disposition: Shutdowns. Consequência:Evasão escolar como uma reação positiva.

**É de suma importância ressaltar que este trabalho é centrado no estudo da reação negativa da disposition,como vemos em massacres em escolas com assassinato de professores e estudantes.**

Num trabalho posterior pensa-se em aperfeiçoar ainda mais a ontologia especificando quem são os tipos de agentes não deficientes que geram a ação mostrando a qual grupo ou colletive pertencem,e do mesmo modo ser feita uma análise mais profunda na linguística do módulo ação de tal maneira que o mode discurso indireto provocativo poderá ser dividido mais claramente de forma taxonômica em rótulos de ofensivo e não ofensivo,e dentro de ofensivo se especializar em bad setences que servirão de dados para alimentar a Machine learning,porém neste momento foca-se de forma mais geral na descoberta das origens que impedem o aluno deficiente de estudar e o que leva a sua evasão escolar na busca de se prevenir a ocorrência ou descobrir os fatos desencadearam uma reação de disposition mais violenta para reabilitação do aluno acometido de transtornos mentais permanentes e prevenir ocorrências em outros alunos.

## 4. Discussões e proposta

Primeiramente ,antes de expor o modelo de proposta,um fato de suma importância a enfatizar neste trabalho é que o uso desta ontologia baseia-se na necessidade de que os alunos com transtornos mentais sejam diagnosticados ,cabendo este papel a instituição de ensino que está também presente na vida social do aluno em grande parte de sua vida e desta forma NÃO devem ser invisibilizados para sua própria proteção, evitando que surja uma confusão mental não intencional provocada pelos outros alunos e professores que em sua ignorância, como a da sua família, foram manipulados a ativar uma disposition de Agressividade física .A agressividade física traz muitos prejuízos sociais tanto para o aluno como a instituição de ensino podendo -se chegar ao extremo de mortes em escolas que poderiam ter sido evitadas como vimos nos casos de tiroteios em instituições de ensino noticiados pelas emissoras de televisão.

4.1. Soluções para identificação dos autores

Diante destes fatos propõe-se a aplicação de um modelo-base no contexto de análise de áudio em ambiente escolar conforme esquema abaixo:

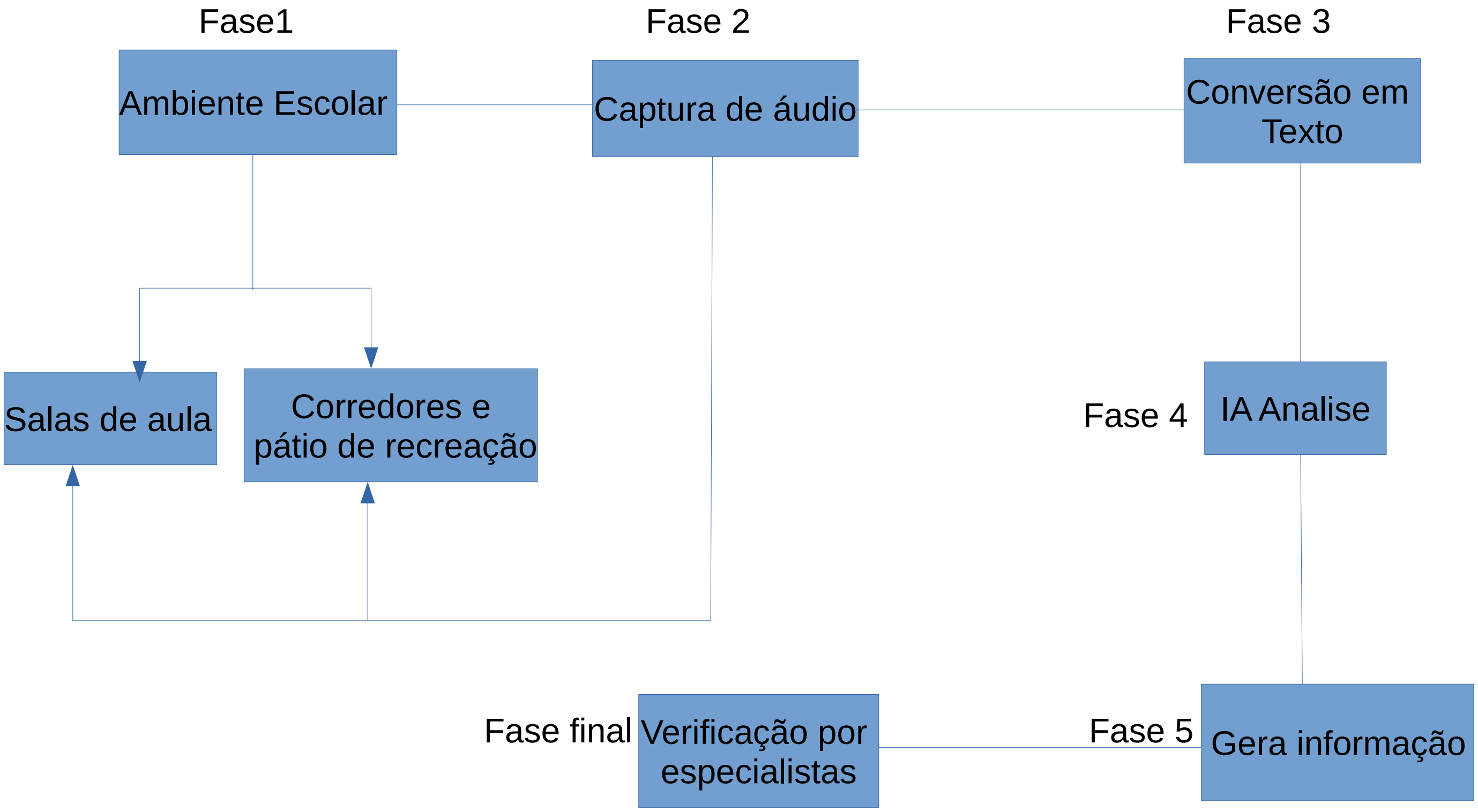

A fase 1 corresponde a definição do domínio da ontologia que já foi bem especificado no início do artigo;na fase 2 temos a obtenção dos dados de voz através de microfones estrategicamente colocados em lugares específicos na fase 1:salas de aula,corredores e pátio, os quais irão gerar arquivos de áudio diários que serão armazenados num servidor; na fase 3 temos a transcrição destes dados podendo utilizar sitio google tape(disponível em https://goodtape.io/gad_source=1&gclid=CjwKCAiA98WrBhAYEiwA2WvhOgb4b6rCD28J-mKwf05-c40cj7XFklxwzd5s5FwEC-h_M7toeP1nzxoCKTAQAvD_BwE) e com criação de um algoritmo para agilizar este procedimento colocando -o como um arquivo de lotes de tal maneira que quando o arquivo de áudio completar 30MB será disparado o envio para o site de conversão de texto do google tape,enviando assim por partes ,colocando o texto numa pasta nomeada com a data do dia da gravação e o arquivo de texto rotulado com a hora da gravação. A fase mais importante focada de forma detalhada é a Análise das sentenças/palavras gatilhos hostis pelo **processamento de linguagem natural** usando a Inteligência Artificial , onde será usado um método baseado no artigo de (Markov et al,2023) que consiste inicialmente em 3 estágios de forma na seleção de dados e aprendizado ativo para incorporar os dados produzidos ao conjunto de aprendizagem para garantir que o sistema de detecção funcione bem no contexto da ontologia.

> “No primeiro estágio um grande volume de dados de produção é selecionado aleatoriamente. Qualquer informação de identificação pessoal (PII) potencial é mascarada. O modelo de moderação mais recente é usado para pontuar essas amostras e descobrir quais delas podem acionar o tipo de discurso indireto provocativo no aluno.
> No segundo estágio, executa-se uma estratégia simples de aprendizagem ativa para selecionar um subconjunto de amostras mais valiosas a serem rotuladas a partir das amostras aleatórias extraídas no primeiro estágio. A estratégia de aprendizagem ativa é composta por três pipelines paralelos. O primeiro baseia-se numa amostragem aleatória, de modo que alguma fração dos dados permaneça consistente com a distribuição de dados subjacente na produção. O segundo seleciona aleatoriamente amostras com pontuação de modelo acima de um determinado limite para cada categoria para identificar prováveis pontos de dados indesejados. O último pipeline adota um conjunto de estratégias de amostragem de incerteza (Lewis e Gale 1994; Lewis e Catlett 1994) para capturar amostras sobre as quais o modelo é mais incerto, onde a pontuação do modelo para essa categoria é mais próxima de 0,5.
> Durante a fase final, todas as amostras selecionadas pelas diferentes estratégias de aprendizagem ativa são agregadas e reponderadas com base nas estatísticas de determinados metadados a elas associados. O peso amostral é configurado para ser proporcional à raiz quadrada da contagem da amostra. Isto ajuda a melhorar a diversidade das amostras selecionadas no que diz respeito aos metadados associados. Atualiza-se a combinação de subestratégias ao longo do tempo com base nas mudanças na distribuição dos dados e nas categorias que mais queremos melhorar em diferentes estágios” (Markov et e al. 2023)

Na fase 5 temos esta informação gerada pela inteligencia artificial que será verificada na realizarão auditorias necessárias,regulares e contínuas para garantir que os dados rotulados continuam a ter uma qualidade suficientemente elevada. A escolha de quais amostras auditar e quais métricas usar para medir a qualidade dos dados é crucial. Voltando a se basear em Markov etc e tal de forma que podemos colocar:

> “A seleção aleatória de alvos de auditoria não pode maximizar o valor da auditoria devido à distribuição desequilibrada entre as categorias. A taxa de concordância verificador auditor (ou seja, precisão) é subo tima porque exemplos indesejados são eventos raros de encontrar e a precisão pode ser arbitrariamente alta devido à abundância de negativos verdadeiros. Em vez

> disso, em cada categoria escolhida, será selecionado aleatoriamente 10 amostras rotuladas como indesejadas e 10 amostras com probabilidade de modelo superior a 50%. Os primeiros ajudam a capturar casos falsos positivos e os últimos fornecem uma estimativa de recall. Em seguida, calcula-se a pontuação F-1 para as amostras escolhidas com base nos rótulos atribuídos pela IA, enquanto usa-se os rótulos atribuídos pelo verificador auditor como verdade fundamental. Este procedimento tem um desempenho muito melhor na prática quando certas categorias de pontos de dados indesejados são raras. A separação das métricas por categoria facilita o reconhecimento de problemas específicos da categoria e o retreinamento adequado da IA de acordo."
>
> (Markov et e al. 2023)

Na fase final coloca -se a nesses cidade de uma equipe interdisciplinar de especialistas auditores que poderá contar com um psicólogo,psiquiatra,pedagogo,diretor da escola e membros do núcleo de atendimento as pessoas com necessidades especiais além do próprio aluno deficiente que saberá quais palavras o atingem e poderá indicar quem as proferiu.

É imprescindível que a equipe multidisciplinar tenha o histórico escolar e familiar também para que tenham uma noção de quais bad setences referenciam indiretamente o aluno com transtornos mentais e a quais fases de vida estão se referindo. O preconceito deve ser banido pelos agentes de ensino pois devem entender se o que estão usando como **fonte de informação** para gerar as agressões verbais não são tautologias ou ,se verdade, em que nível de consciência e noção de realidade o aluno deficiente tinha na época para ter tomado atitudes não padrão e que,principalmente independente de ser verdade ou não, não devem ser usadas em ambientes de ensino como discurso de calunia, injuria ,difamação ou ódio .

## 5.Conclusão

De acordo com artigo da CNN no Brasil houve pelo menos 16 ataques com alunos armados em escolas nos últimos vinte anos. Em nosso Estado tivemos ao menos em 2 escolas relatadas somente em Aracruz , em Novembro de 2022.Em coletiva de imprensa o governador Casa Grande disse que o autor dos ataques confessou ter cometido o crime e não falou sobre sua motivação,mas disse ter planejado o ataque por cerca de 2 anos.

Pessoas não nascem monstruosas, elas são "adestradas" para se tornarem assim. Esta ontologia é só o principio,baseado num relato pessoal da autora, para mostrar que a agressividade não surge ao acaso e sim tem uma causa ambiental,pessoal, e que depende de uma personalidade. Em microambientes, com maior facilidade de investigação por ser uma área pequena,fica mais fácil descobrir de onde vem as origens para saber o que fazer e tomar atitudes para impedir que ocorra um desastre social:A vida real torna-se um laboratório para mostrar o que o ser humano é capaz de fazer dotado de uma irracionalidade em sua pseudo liberdade ilusória de liberdade de expressão que se transforma em calúnia,injúria e difamação,um ódio intrínseco, pois está cego demais para perceber que apenas está segregando para favorecer alguns grupos mais apropriados aos seus anseios de momento ou futuros.

Espera-se que a partir desta iniciativa surjam outros modelos UFO, especificando mais os agentes, as intenções, acões e as dispositions, com ajuda de especialistas de cada área,ou o contrário, que a sociedade ignore pois a pretensão é que sempre obscureçam pessoas assim,em prol de um sistema econômico que sempre favoreceu sempre favorecerá alguns considerados de supremacia intelectual e financeira em prol da extinção de uns “desajustados” que não se adéquam e sonham em transformar o mundo em uma realidade mais justa. Os dados agora serão jogados, literalmente: Que corrijam o que considerarem errado e então veremos quem realmente está com a razão.

REFERENCIAS